\documentclass[preprint, aps, prd]{revtex4-1}

\def\pd{\partial}
\def\mc{\mathcal}
\def\ul{\underline}

\usepackage[dvips]{graphicx}
\usepackage{amssymb}
\usepackage[bbgreekl]{mathbbol}
\usepackage{amssymb,amsmath}
\usepackage{graphicx}
\usepackage{dsfont}
\usepackage{caption}
\usepackage{subcaption}
\usepackage{mathtools}
\usepackage{verbatim}
\usepackage{graphicx}
\usepackage{multirow}
\usepackage[outline]{contour}
\usepackage{xcolor,colortbl}
\makeatother
\begin{document}

\title{Janus and RG-flow interfaces from matter-coupled $F(4)$ gauged supergravity}

\author{Parinya Karndumri} \email[REVTeX Support:
]{parinya.ka@hotmail.com} \affiliation{String Theory and
Supergravity Group, Department of Physics, Faculty of Science,
Chulalongkorn University, 254 Phayathai Road, Pathumwan, Bangkok
10330, Thailand}

\date{\today}
\begin{abstract}
We study supersymmetric Janus solutions from matter-coupled $F(4)$ gauged supergravity coupled to three vector multiplets and $SO(4)\sim SO(3)\times SO(3)$ gauge group. There are two supersymmetric $AdS_6$ vacua preserving all supersymmetries with $SO(3)\times SO(3)$ and $SO(3)_{\textrm{diag}}$ symmetries dual to $N=2$ SCFTs in five dimensions. We consider a truncation to $SO(2)_{\textrm{diag}}\subset SO(3)_{\textrm{diag}}$ singlet scalars and find a number of new supersymmetric Janus solutions preserving eight supercharges. These solutions holographically describe conformal interfaces within $N=2$ five-dimensional SCFTs involving deformations by source terms and vacuum expectation values of relevant and irrelevant operators. Apart from the Janus solutions interpolating between $SO(3)\times SO(3)$ $AdS_6$ vacua, some of the solutions have $SO(3)_{\textrm{diag}}$ $AdS_6$ vacua generated by holographic RG flows from the $SO(3)\times SO(3)$ phases on both sides. We also give an evidence for solutions describing RG-flow interfaces with $SO(3)\times SO(3)$ $AdS_6$ vacuum on one side and $SO(3)_{\textrm{diag}}$ $AdS_6$ vacuum on the other side. The solutions provide first examples of Janus solutions involving more than one $AdS_6$ vacuum in six-dimensional gauged supergravity.
\end{abstract}
\maketitle
\section{Introduction}
The study of Janus solutions leads to holographic descriptions of conformal interfaces within superconformal field theories (SCFTs). The solutions can be obtained directly from string/M-theory or by uplifting $AdS_d$-sliced domain wall solutions from $(d+1)$-dimensional gauged supergravity. Since the first Janus solution found in \cite{Bak_Janus}, a number of previous works studying Janus solutions from both approaches have appeared even though, in some cases, the embedding to higher-dimensions is presently unknown \cite{Freedman_Janus}-\cite{Guarino_S_fold_Janus}. It is well known that finding solutions from lower-dimensional gauged supergravity and subsequently uplifting to ten or eleven dimensions are much simpler than constructing the corresponding solutions in higher dimensions. Various holographic computations are also more traceable in gauged supergravity. Although a number of gauged supergravities currently have no known higher dimensional origins, the study of Janus solutions within the framework of lower-dimensional gauged supergravity is, nevertheless, still useful and could give some insights to the nature of conformal interfaces. 
\\
\indent In six dimensions, only the half-maximal $N=(1,1)$ gauged supergravity coupled to vector multiplets can lead to supersymmetric $AdS_6$ vacua in agreement with the classification of superconformal algebra in five dimensions \cite{Nahm_res}. Pure $N=(1,1)$, usually called $F(4)$, gauged supergravity has been constructed long ago in \cite{F4_Romans}. The matter coupled $F(4)$ gauged supergravity has been constructed more recently in \cite{F4SUGRA1} and \cite{F4SUGRA2}. Since the study of $AdS_6$ vacua and holographic RG flows initiated in \cite{F4_flow} and \cite{5DSYM_from_F4}, various types of holographic solutions have been found from this gauged supergravity \cite{6D_Janus,6D_twist,AdS6_BH_Minwoo1,AdS6_BH_Zaffaroni,AdS6_BH_Minwoo,AdS6_BH} including the general conditions on the existence of supersymmetric $AdS_6$ vacua given in \cite{AdS6_Jan}. In particular, the Janus solution found in \cite{6D_Janus} is the first example of Janus solutions within a six-dimensional framework. The $F(4)$ gauged supergravity considered in \cite{6D_Janus} couples to a single vector multiplet and has $SU(2)\sim SO(3)$ gauge group as in the pure $F(4)$ gauged supergravity. This gauged supergravity admits only one supersymmetric $AdS_6$ vacuum, so the Janus solution found in \cite{6D_Janus} necessarily interpolates between this $AdS_6$ vacuum on each side of the interface.     
\\
\indent We will look for more interesting supersymmetric Janus solutions involving more than one supersymmetric $AdS_6$ vacuum. To this end, we consider a larger $SO(4)\sim SO(3)\times SO(3)$ gauge group in which two supersymmetric $N=(1,1)$ $AdS_6$ vacua with $SO(4)$ and $SO(3)_{\textrm{diag}}\subset SO(3)\times SO(3)$ symmetries are known to exist \cite{F4_flow}. This gauged supergravity is obtained from $F(4)$ gauged supergravity coupled to three vector multiplets. We will work within a consistent truncation consisting of the metric and scalars which are singlet under $SO(2)_{\textrm{diag}}\subset SO(3)_{\textrm{diag}}$. This sector has also been considered recently in \cite{AdS6_BH} in order to find supersymmetric $AdS_6$ black hole solutions. We will point out that a simpler truncation with $SO(3)_{\textrm{diag}}$ residual symmetry, the minimal scalar sector that can accommodate both of the aforementioned supersymmetric $AdS_6$ vacua considered in \cite{F4_flow}, does not admit Janus solutions. This means that the $SO(3)_{\textrm{diag}}$ invariant sector can only support the flat domain walls studied in \cite{F4_flow}. It is then natural to look for Janus solutions in a larger truncation. We will see that $SO(2)_{\textrm{diag}}$ sector can indeed lead to new supersymmetric Janus solutions interpolating among the two supersymmetric $AdS_6$ vacua. It should also be pointed out that the six-dimensional gauged supergravity considered here has currently no known higher-dimensional origin, so there is no clear holographic description within the framework of string/M-theory up to now. However, given the unique role of the matter-coupled $F(4)$ gauged supergravity and $N=2$ SCFTs in five dimensions, the solutions given here could still be useful in the holographic study of conformal interfaces within five-dimensional $N=2$ SCFTs.   
\\
\indent The paper is organized as follows. After a brief review of the $F(4)$ gauged supergravity coupled to three vector multiplets with $SO(4)\sim SO(3)\times SO(3)$ gauge group in section \ref{6D_SO4gaugedN2}, we consider the metric ansatz in the form of an $AdS_5$-sliced domain wall within the truncation to $SO(2)_{\textrm{diag}}$ invariant scalars in section \ref{Janus_solutions}. We analysize the BPS conditions and derive a relevant set of BPS equations and subsequently give various types of numerical Janus solutions. We end the paper with some conclusions and comments in section \ref{conclusion}. In the appendix, we give the explicit form of the scalar potential and the bosonic field equations within the $SO(2)_{\textrm{diag}}$ truncation considered in the paper. 

\section{Matter-coupled $F(4)$ gauged supergravity in six dimensions}\label{6D_SO4gaugedN2}
We first give a review of matter-coupled $F(4)$ gauged supergravity in six dimensions constructed in \cite{F4SUGRA1} and \cite{F4SUGRA2}. We mainly focus on the gauged supergravity with $SO(4)\sim SO(3)\times SO(3)$ gauge group obtained by gauging the half-maximal $N=(1,1)$ supergravity coupled to three vector multiplets. We closely follow the conventions of \cite{F4SUGRA1} and \cite{F4SUGRA2} with the metric signature $(-+++++)$. Since this gauged supergravity has already been studied in a number of previous works, we will only collect relevant formulae for constructing supersymmetric Janus solutions with only the metric and scalar fields non-vanishing.
\\
\indent The matter-coupled $F(4)$ gauged supergravity with $SO(3)\times SO(3)$ gauge group is obtained by gauging the half-maximal $N=(1,1)$ supergravity coupled to three vector multiplets. The gravity and three vector multiplets consist of the following component fields 
\begin{equation}
\left(e^{\hat{\mu}}_\mu,\psi^A_\mu, A^\alpha_\mu, B_{\mu\nu}, \chi^A,
\sigma\right)\qquad \textrm{and}\qquad (A_\mu,\lambda_A,\phi^\alpha)^I, \qquad I=1, 2, 3\, .
\end{equation}
The conventions for various indices are as follows. Space-time and tangent space or flat indices are denoted respectively by $\mu,\nu=0,\ldots ,5$ and $\hat{\mu},\hat{\nu}=0,\ldots, 5$. Indices $A,B,\ldots =1,2$ correspond to the fundamental representation of $SU(2)_R\sim USp(2)_R\sim SO(3)_R$ R-symmetry while indices $\alpha,\beta=(0,r)$ will be split into singlet, $0$, and adjoint indices, $r,s,\ldots =1,2,3$, of $SU(2)_R$. 
\\
\indent The bosonic fields are given by the graviton $e^{\hat{\mu}}_\mu$, a two-form field $B_{\mu\nu}$, seven vector fields $A^\Lambda=(A^\alpha_\mu,A^I_\mu)$, $\Lambda=(\alpha,I)=0,1,2,\dots, 6$, and $13$ scalars described by the scalar manifold 
\begin{equation}
\mathbb{R}^+\times SO(4,3)/SO(4)\times SO(3)\, .
\end{equation}
The $\mathbb{R}^+$ factor corresponds to the dilaton $\sigma$ while the $12$ scalars from the vector multiplets $\phi^{\alpha I}$ are encoded in the $SO(4,3)/SO(4)\times SO(3)$ coset representative
\begin{equation}
{L^\Lambda}_{\ul{\Sigma}}=({L^\Lambda}_\alpha,{L^\Lambda}_I).
\end{equation}
The inverse of ${L^\Lambda}_{\ul{\Sigma}}$ will be denoted by ${(L^{-1})^{\ul{\Lambda}}}_\Sigma=({(L^{-1})^{\alpha}}_\Sigma,{(L^{-1})^{I}}_\Sigma)$. Finally, the fermionic fields are given by two gravitini $\psi^A_\mu$, two spin-$\frac{1}{2}$ fields $\chi^A$, and six gaugini $\lambda^I_A$. 
\\
\indent With only the metric and scalars non-vanishing, the bosonic Lagrangian for the resulting gauged supergravity can be written as 
\begin{eqnarray}
e^{-1}\mathcal{L}&=&\frac{1}{4}R-\pd_\mu \sigma\pd^\mu \sigma
-\frac{1}{4}P^{I\alpha}_\mu P^{\mu}_{I\alpha}-V\label{Lar}
\end{eqnarray}
with $e=\sqrt{-g}$. The vielbein on $SO(4,3)/SO(4)\times SO(3)$, denoted by $P^{I\alpha}_\mu$, can be obtained from the left-invariant 1-form
\begin{equation}
{\Omega^{\ul{\Lambda}}}_{\ul{\Sigma}}=
{(L^{-1})^{\ul{\Lambda}}}_{\Pi}\nabla {L^\Pi}_{\ul{\Sigma}}
\qquad \textrm{with}\qquad \nabla
{L^\Lambda}_{\ul{\Sigma}}={dL^\Lambda}_{\ul{\Sigma}}
-f^{\phantom{\Gamma}\Lambda}_{\Gamma\phantom{\Lambda}\Pi}A^\Gamma
{L^\Pi}_{\ul{\Sigma}}
\end{equation}
by the following identification  
\begin{equation}
P^I_{\phantom{s}\alpha}=(P^I_{\phantom{a}0},P^I_{\phantom{a}r})=(\Omega^I_{\phantom{a}0},\Omega^I_{\phantom{a}r}).
\end{equation}
The remaining components $(\Omega^{rs},\Omega^{r0},\Omega^{IJ})$ are the $SO(4)\times SO(3)$ composite connections, and ${f^\Lambda}_{\Pi\Gamma}$ are structure constants of the gauge group. We also note that all $SO(4,3)$ indices $\Lambda,\Sigma,\ldots$ are raised and lowered by the $SO(4,3)$ invariant tensor
\begin{equation}
\eta^{\Lambda\Sigma}=\eta_{\Lambda\Sigma}=(\delta_{\alpha\beta},-\delta_{IJ}).
\end{equation}
\indent The scalar potential is given by
\begin{eqnarray}
V&=&-e^{2\sigma}\left[\frac{1}{36}A^2+\frac{1}{4}B^iB_i+\frac{1}{4}\left(C^I_{\phantom{s}t}C_{It}+4D^I_{\phantom{s}t}D_{It}\right)\right]
+m^2e^{-6\sigma}\mc{N}_{00}\nonumber \\
& &-me^{-2\sigma}\left[\frac{2}{3}AL_{00}-2B^iL_{0i}\right]
\end{eqnarray}
with $\mc{N}_{00}$ being the $00$ component of $\mc{N}_{\Lambda\Sigma}$ defined as
\begin{eqnarray}
\mc{N}_{\Lambda\Sigma}=L_{\Lambda\alpha}{(L^{-1})^\alpha}_\Sigma-L_{\Lambda I}{(L^{-1})^I}_\Sigma=(\eta L L^T\eta)_{\Lambda\Sigma}\, .
\end{eqnarray}
\indent In the present paper, we are only interested in compact gauge groups that lead to supersymmetric $AdS_6$ vacua. According to the general result of \cite{AdS6_Jan}, a compact gauge group of this type must take the form $SO(3)\times H_n\subset SO(4)\times SO(n)\subset SO(4,n)$ with $SO(3)$ being a diagonal subgroup of $SO(3)\times SO(3)\sim SO(4)$ and $H_n$ being an $n$-dimensional compact subgroup of $SO(n)$. For the case of $n=3$ vector multiplets under consideration here, we have $H_3=SO(3)$ resulting in $SO(3)\times SO(3)\sim SU(2)\times SU(2)$ gauge group. The most general embedding of this gauge group in the $SO(4,3)$ global symmetry is then given by the following components of the structure constants 
\begin{equation}
{f^\Lambda}_{\Pi\Sigma}=(g_1\epsilon_{rst},g_2\epsilon_{IJK})
\end{equation}
with $g_1$ and $g_2$ being the corresponding gauge coupling constants for the two $SO(3)$ factors. The first $SO(3)$ is identified with the R-symmetry gauged by $A^r_\mu$ while the second factor is gauged by $A^I_\mu$ from the three vector multiplets. The fermion-shift matrices, appearing in the scalar potential and also in the supersymmetry transformations of fermions, are defined as follows
\begin{eqnarray}
A&=&\epsilon^{rst}K_{rst},\qquad B^i=\epsilon^{ijk}K_{jk0},\\
{C_I}^t&=&\epsilon^{trs}K_{rIs},\qquad D_{It}=K_{0It}
\end{eqnarray}
where
\begin{eqnarray}
K_{rst}&=&g_1\epsilon_{lmn}L^l_{\phantom{r}r}(L^{-1})_s^{\phantom{s}m}L_{\phantom{s}t}^n+
g_2\epsilon_{IJK}L^I_{\phantom{r}r}(L^{-1})_s^{\phantom{s}J}L_{\phantom{s}t}^K,\nonumber
\\
K_{rs0}&=&g_1\epsilon_{lmn}L^l_{\phantom{r}r}(L^{-1})_s^{\phantom{s}m}L_{\phantom{s}0}^n+
g_2\epsilon_{IJK}L^I_{\phantom{r}r}(L^{-1})_s^{\phantom{s}J}L_{\phantom{s}0}^K,\nonumber
\\
K_{rIt}&=&g_1\epsilon_{lmn}L^l_{\phantom{r}r}(L^{-1})_I^{\phantom{s}m}L_{\phantom{s}t}^n+
g_2\epsilon_{IJK}L^I_{\phantom{r}r}(L^{-1})_I^{\phantom{s}J}L_{\phantom{s}t}^K,\nonumber
\\
K_{0It}&=&g_1\epsilon_{lmn}L^l_{\phantom{r}0}(L^{-1})_I^{\phantom{s}m}L_{\phantom{s}t}^n+
g_2\epsilon_{IJK}L^I_{\phantom{r}0}(L^{-1})_I^{\phantom{s}J}L_{\phantom{s}t}^K\, .
\end{eqnarray}
\indent Supersymmetry transformations for the fermionic fields are given by
\begin{eqnarray}
\delta\psi_{\mu
A}&=&D_\mu\epsilon_A-\frac{1}{24}\left(Ae^\sigma+6me^{-3\sigma}(L^{-1})_{00}\right)\epsilon_{AB}\gamma_\mu\epsilon^B\nonumber
\\
& &-\frac{1}{8}
\left(B_te^\sigma-2me^{-3\sigma}(L^{-1})_{t0}\right)\gamma^7\sigma^t_{AB}\gamma_\mu\epsilon^B,\label{delta_psi}\\
\delta\chi_A&=&\frac{1}{2}\gamma^\mu\pd_\mu\sigma\epsilon_{AB}\epsilon^B+\frac{1}{24}
\left[Ae^\sigma-18me^{-3\sigma}(L^{-1})_{00}\right]\epsilon_{AB}\epsilon^B\nonumber
\\
& &-\frac{1}{8}
\left[B_te^\sigma+6me^{-3\sigma}(L^{-1})_{t0}\right]\gamma^7\sigma^t_{AB}\epsilon^B,\label{delta_chi}\\
\delta
\lambda^{I}_A&=&P^I_{ri}\gamma^\mu\pd_\mu\phi^i\sigma^{r}_{\phantom{s}AB}\epsilon^B+P^I_{0i}
\gamma^7\gamma^\mu\pd_\mu\phi^i\epsilon_{AB}\epsilon^B-\left(2i\gamma^7D^I_{\phantom{s}t}+C^I_{\phantom{s}t}\right)
e^\sigma\sigma^t_{AB}\epsilon^B \nonumber
\\
& &+2me^{-3\sigma}(L^{-1})^I_{\phantom{ss}0}
\gamma^7\epsilon_{AB}\epsilon^B\label{delta_lambda}
\end{eqnarray}
with ${\sigma^{rA}}_B$ being Pauli matrices and $\epsilon_{AB}=-\epsilon_{BA}$. The $SU(2)$ fundamental indices $A,B,\ldots$ can be raised and lowered by $\epsilon^{AB}$ and $\epsilon_{AB}$ with the convention $T^A=\epsilon^{AB}T_B$ and $T_A=T^B\epsilon_{BA}$. The covariant derivative of $\epsilon_A$ is
given by
\begin{equation}
D_\mu \epsilon_A=\pd_\mu
\epsilon_A+\frac{1}{4}\omega_\mu^{ab}\gamma_{ab}\epsilon_A+\frac{i}{2}\sigma^r_{AB}
\left[\frac{1}{2}\epsilon^{rst}\Omega_{\mu st}-i\gamma_7
\Omega_{\mu r0}\right]\epsilon^B\, .
\end{equation}
$\gamma^a$ matrices satisfy the Clifford algebra
\begin{equation}
\{\gamma^a,\gamma^b\}=2\eta^{ab},\qquad
\eta^{ab}=\textrm{diag}(-1,1,1,1,1,1),
\end{equation}
with the chirality matrix defined by $\gamma_7=i\gamma^0\gamma^1\gamma^2\gamma^3\gamma^4\gamma^5$ satisfying
$\gamma_7^2=-\mathbf{1}$. 
\section{Supersymmetric Janus solutions}\label{Janus_solutions}
In this section, we consider a particular truncation of the matter-coupled $F(4)$ gauged supergravity with $SO(3)\times SO(3)$ gauge group described in the previous section. This truncation contains all scalars which are singlet under $SO(2)_{\textrm{diag}}\subset SO(3)_{\textrm{diag}}\subset SO(3)\times SO(3)$. We then consider supersymmetric Janus solutions obtained by solving the BPS equations which in turn arise from supersymmetry transformations of fermionic fields.   

\subsection{$SO(2)_{\textrm{diag}}$ singlet sector and supersymmetric $AdS_6$ vacua}
Including the dilaton which is a singlet under the full gauge group, there are five $SO(2)_{\textrm{diag}}$ singlet scalars. Since this sector has already been considered recently in \cite{AdS6_BH}, we will only repeat relevant formulae here and refer to \cite{AdS6_BH} for more detail. 
\\
\indent In terms of the $GL(7,\mathbb{R})$ matrices
\begin{equation}
(e^{\Lambda \Sigma})_{\Gamma \Pi}=\delta^\Lambda_{
\Gamma}\delta^\Sigma_{\Pi},\qquad \Lambda, \Sigma,\Gamma,
\Pi=0,\ldots ,6,
\end{equation}
the compact $SO(4)\times SO(3)$ generators are given by
\begin{eqnarray}
SO(4):\qquad
J^{\alpha\beta}&=&e^{\beta,\alpha}-e^{\alpha,\beta},\qquad \alpha,\beta=0,1,2,3,\nonumber \\
SO(3):\qquad \tilde{J}^{IJ}&=&e^{J+3,I+3}-e^{I+3,J+3},\qquad I,J=1,2,3
\end{eqnarray}
while non-compact generators can be identified with
\begin{eqnarray}
Y_{\alpha I}=e^{\alpha,I+3}+e^{I+3,\alpha}\, .
\end{eqnarray}
The structure constants given in the previous section imply that the $SO(3)\times SO(3)$ gauge generators are given respectively by $J^{rs}$ and $\tilde{J}^{IJ}$. 
\\
\indent With $SO(2)_{\textrm{diag}}$ generated by $J^{12}+\tilde{J}^{12}$, the four singlet scalars from $SO(4,3)/SO(4)\times SO(3)$ can be described by the coset representative
\begin{equation}
L=e^{\phi_0 Y_{03}}e^{\phi_1 (Y_{11}+Y_{22})}e^{\phi_2 Y_{33}}e^{\phi_3 (Y_{12}-Y_{21})}\, .\label{L_SO2diag}
\end{equation}
To consistently truncate out all the gauge fields, we need to set $\phi_3=0$, see \cite{AdS6_BH} for the list of all bosonic field equations. We will accordingly take $\phi_3=0$ from now on. The explicit form of the scalar potential is given in the appendix. This potential admits two supersymmetric $AdS_6$ critical points. The first one is given by
\begin{eqnarray}
& &\phi_0=\phi_1=\phi_2=0,\qquad \sigma=\frac{1}{4}\ln
\left[\frac{3m}{g_1}\right],\nonumber \\
& &V_0=-20m^2\left(\frac{g_1}{3m}\right)^{\frac{3}{2}},\qquad L=\frac{1}{2m}\left(\frac{3m}{g_1}\right)^{\frac{3}{4}}\, .\label{SO4_AdS6}
\end{eqnarray}
$V_0$ is the cosmological constant, and $L$ is the $AdS_6$ radius related to $V_0$ via the following relation
\begin{equation}
L^2=-\frac{5}{V_0}\, .
\end{equation}
In the above equation, we have taken $m>0$ and $g_1>0$. This $AdS_6$ vacuum preserves $N=(1,1)$ supersymmetry and the full $SO(3)\times SO(3)$ gauge symmetry since $\sigma$ is invariant under the $SO(3)\times SO(3)$ gauge group. It should also be noted that we can set $g_1=3m$ to have vanishing dilaton at the critical point. 
\\
\indent The second critical point is given by 
\begin{eqnarray}
& &\phi_0=0,\qquad \phi_1=\phi_2=\frac{1}{2}\ln \left[\frac{g_1+g_2}{g_2-g_1}\right],\qquad
\sigma=\frac{1}{4}\ln
\left[\frac{3m\sqrt{g_2^2-g_1^2}}{g_1g_2}\right],\nonumber \\
& &V_0=-20m^2\left[\frac{g_1g_2}{3m\sqrt{g_2^2-g_1^2}}\right]^{\frac{3}{2}},\qquad L=\frac{1}{2m}\left(\frac{3m\sqrt{g_2^2-g_1^2}}{g_1g_2}\right)^{\frac{3}{4}}\,
.\label{SO3_AdS6}
\end{eqnarray}
This critical point also preserves $N=(1,1)$ supersymmetry but breaks the $SO(3)\times SO(3)$ gauge group to $SO(3)_{\textrm{diag}}$ subgroup. According to the AdS/CFT correspondence, these two $AdS_6$ vacua should be dual to $N=2$ SCFTs in five dimensions, and holographic RG flows interpolating between them have already been studied in detail in \cite{F4_flow}, see also \cite{5DSYM_from_F4} for solutions describing RG flows to non-conformal phases. In the present work, we will consider Janus solutions interpolating among these vacua. 

\subsection{BPS equations for Janus solutions}
As previously mentioned, Janus solutions holographically describe four-dimensional conformal interfaces within five-dimensional $N=2$ SCFTs. We will consider the metric ansatz in the form of an $AdS_5$-sliced domain wall given by
\begin{equation}
ds^2=e^{2f(r)}d\tilde{s}^2+dr^2\label{Janus_ansatz}
\end{equation}  
with $d\tilde{s}^2$ being the metric on a unit $AdS_5$ space. We now solve the BPS conditions arising from setting the supersymmetry transformations of fermionic fields $(\psi^A_\mu,\chi^A,\lambda^{IA})$ to zero. The analysis closely follows the procedures given in \cite{Bobev_5D_Janus2,warner_Janus} and \cite{6D_Janus}. 
\\
\indent We begin with the conditions obtained from the gaugino variations, $\delta\lambda^{IA}=0$. For $I=1,2,3$, these variations lead to the following conditions 
\begin{eqnarray}
\delta\lambda^{1A}&:&\quad \phi_1'{(\sigma_1)^A}_B\gamma_{\hat{r}}\epsilon^B+M_1{(\sigma_1)^A}_B\epsilon^B+i\gamma_7M_2{(\sigma_2)^A}_B\epsilon^B=0,\label{dLambda1}\\
\delta\lambda^{2A}&:&\quad \phi_1'{(\sigma_2)^A}_B\gamma_{\hat{r}}\epsilon^B-i\gamma_7M_2{(\sigma_1)^A}_B\epsilon^B+M_1{(\sigma_2)^A}_B\epsilon^B=0,\label{dLambda2}\\
\delta\lambda^{3A}&:&\quad \left[\phi_2'{(\sigma_3)^A}_B-\cosh\phi_2\phi'_0\delta^A_B\gamma_7\right]\gamma_{\hat{r}}\epsilon^B+M_0\gamma_7\epsilon^A+M_3{(\sigma_3)^A}_B\epsilon^B
=0\quad\label{dLambda3}
\end{eqnarray}
with
\begin{eqnarray}
& &M_0=2me^{-3\sigma}\sinh\phi_0\cosh\phi_2,\nonumber \\
& &M_1=-e^\sigma\sinh2\phi_1(g_1\cosh\phi_2-g_2\cosh\phi_0\sinh\phi_2),\nonumber \\
& &M_2=-g_2e^\sigma\sinh\phi_0\sinh2\phi_1,\nonumber \\
& &M_3=g_2e^\sigma\cosh\phi_0\cosh\phi_2(\cosh2\phi_1-1)-g_1e^\sigma\sinh\phi_2(\cosh2\phi_1+1).
\end{eqnarray}
For convenience, we have raised the $SU(2)$ index $A$ in the above equations.
\\
\indent By multiplying equation \eqref{dLambda1} by $\sigma_1$ or multiplying equation \eqref{dLambda2} by $\sigma_2$, we obtain the same condition of the form
\begin{equation}
-\phi'_1\gamma_{\hat{r}}\epsilon^A=M_1\epsilon^A-M_2\gamma_7{(\sigma_3)^A}_B\epsilon^B\, .\label{gamma_r_proj}
\end{equation}
This is a projection condition in which consistency requires
\begin{equation}
{\phi'_1}^2=M_1^2+M_2^2\, .\label{phi1_eq}
\end{equation}
Multiply equation \eqref{dLambda3} by $-\phi'_1\gamma_{\hat{r}}$ and use the condition \eqref{gamma_r_proj}, we find
\begin{eqnarray}
\cosh\phi_2\phi'_1\phi'_0&=&-M_0M_1-M_2M_3,\label{phi0_eq}\\
\phi'_1\phi'_2&=&M_1M_3-M_0M_2\, .\label{phi2_eq}
\end{eqnarray}
These equations arise from the requirement that we do not want to impose another projector on $\epsilon^A$, so the $\epsilon^A$ and $\gamma_7\epsilon^A$ terms must vanish separately. 
\\
\indent We now move to $\delta \chi^A=0$ condition which is given by
\begin{equation}
-\frac{1}{2}\sigma'\gamma_{\hat{r}}\epsilon^A=N_0\epsilon^A+N_3\gamma_7{(\sigma_3)^A}_B\epsilon^B\label{dchi_eq}
\end{equation}
with
\begin{eqnarray}
N_0&=&\frac{1}{8}e^{-3\sigma}\cosh\phi_0\left[6m+g_2e^{4\sigma}\sinh\phi_2(\cosh2\phi_1-1)\right]\nonumber \\
& &-\frac{1}{8}g_1e^\sigma\cosh\phi_2(\cosh2\phi_1+1),\nonumber \\
N_3&=&\frac{1}{8}e^{-3\sigma}\sinh\phi_0\left[e^{4\sigma}g_2(\cosh2\phi_1-1)-6m\sinh\phi_2\right].
\end{eqnarray}
We again multiply equation \eqref{dchi_eq} by $-\phi'_1\gamma_{\hat{r}}$ and take into account the projector \eqref{gamma_r_proj}. This leads to the flow equation for the dilaton from the $\epsilon^A$ terms 
\begin{equation}
\sigma'\phi'_1=2(N_0M_1-N_3M_2)\label{sigma_eq}
\end{equation}
together with an algebraic constraint from the $\gamma_7\epsilon^A$ terms
\begin{equation}
M_1N_3=-N_0M_2\, .\label{constraint1}
\end{equation}
Explicitly, this algebraic constraint takes the form of
\begin{equation}
g_1g_2e^{4\sigma}=3mg_2\cosh\phi_0\cosh\phi_2-3mg_1\sinh\phi_2
\end{equation}
which imposes a relation among the scalar fields. It can be straightforwardly check that this constraint is indeed consistent with the flow equations given in \eqref{phi1_eq}, \eqref{phi0_eq}, \eqref{phi2_eq}, and \eqref{sigma_eq}.
\\
\indent We finally consider the gravitino variations which are given by
\begin{equation}
D_\mu\epsilon^A=S_0\gamma_\mu \epsilon^A+S_3{(\sigma_3)^A}_B\gamma_7\gamma_\mu \epsilon^B
\end{equation}
for 
\begin{eqnarray}
S_0&=&-\frac{1}{8}g_1e^\sigma\cosh\phi_2(\cosh2\phi_1+1)\nonumber \\
& &-\frac{1}{8}e^{-3\sigma}\cosh\phi_0\left[2m-g_2e^{4\sigma}\sinh\phi_2(\cosh2\phi_1-1)\right],\nonumber \\
S_3&=&-\frac{1}{8}e^{-3\sigma}\sinh\phi_0\left[g_2e^{4\sigma}(\cosh2\phi_1-1)+2m\sinh\phi_2\right].
\end{eqnarray}
With the six-dimensional coordinates split as $x^\mu=(x^m,r)$, the variation along the $AdS_5$ coordinates $x^m$ takes the form
\begin{equation}
D_m\epsilon=(S_0+S_3\gamma_7\sigma_3)\gamma_m\epsilon
\end{equation}
in which we have omitted the R-symmetry indices for convenience. Using the integrability condition
\begin{equation}
\left[D_m,D_n\right]\epsilon=\frac{1}{4}R_{mnpq}\gamma^{pq}\epsilon
\end{equation}
together with the Riemann tensor computed from the metric \eqref{Janus_ansatz}
\begin{equation}
R_{mnpq}=-\left({f'}^2+e^{-2f}\right)(g_{mp}g_{nq}-g_{mq}g_{np}),
\end{equation}
we arrive at the following equation
\begin{equation}
{f'}^2+e^{-2f}=4(S_0^2+S_3^2).\label{A_eq1}
\end{equation}
Following \cite{6D_Janus}, we can also rewrite the covariant derivative along $AdS_5$ as
\begin{equation}
D_m\epsilon=\widetilde{D}_m\epsilon+\frac{1}{2}f'e^f\tilde{\gamma}_m\gamma_r=e^f(S_0+S_3\gamma_7\sigma_3)\tilde{\gamma}_m\epsilon
\end{equation} 
with $\widetilde{D}_m$ being the covariant derivative along a unit $AdS_5$. 
\\
\indent Using the projection \eqref{gamma_r_proj}, we find
\begin{equation}
e^{-f}\widetilde{D}_m\epsilon=\tilde{\gamma}_m\left[\frac{1}{2}\frac{f'}{\phi'_1}M_1+S_0-\left(\frac{1}{2}\frac{f'}{\phi'_1}M_2+S_3\right)\gamma_7\sigma_3\right]\epsilon\, .
 \end{equation} 
Using the integrability condition on a unit $AdS_5$ 
\begin{equation}
\left[\widetilde{D}_m,\widetilde{D}_n\right]\epsilon=\frac{1}{4}\tilde{R}_{mnpq}\tilde{\gamma}^{pq}\epsilon
\end{equation}
for 
\begin{equation}
\tilde{R}_{mnpq}=-(\tilde{g}_{mp}\tilde{g}_{nq}-\tilde{g}_{mq}\tilde{g}_{np}),
\end{equation}
we find
\begin{equation}
e^{-2f}={f'}^2+4\frac{f'}{\phi'_1}(M_1S_0+M_2S_3)+4(S_0^2+S_3^2).\label{AdS5_integrability}
\end{equation}
Using $e^{-2f}$ from \eqref{A_eq1}, we find another form of the BPS equations for $f'$ as follows
\begin{equation}
f'=-\frac{2}{\phi'_1}(S_0M_1+S_3M_2).\label{A_eq2}
\end{equation}
On the other hand, substituting ${f'}^2$ from \eqref{A_eq1} in equation \eqref{AdS5_integrability}, we obtain another algebraic constraint
\begin{equation}
e^{-2f}=4(S_0^2+S_3^2)-\frac{4(S_0M_1+S_3M_2)^2}{M_1^2+M_2^2}\label{constraint2}
\end{equation} 
after using ${\phi'_1}^2$ from \eqref{phi1_eq}. 
\\
\indent It can be verified that this constraint is compatible with all the previously derived BPS equations. Furthermore, we have also explicitly checked that the resulting BPS equations and the two algebraic constraints \eqref{constraint1} and \eqref{constraint2} indeed satisfy the second-ordered field equations given in the appendix. From equation \eqref{constraint2}, we also see that when $\phi_0=0$, the solutions are necessarily flat domain walls. This can be seen as follows. For $\phi_0=0$, we find $S_3=M_2=0$ which gives $e^{-2f}=0$. This has also been pointed out in \cite{6D_Janus} in a simpler truncation of $F(4)$ gauged supergravity. Moreover, this rules out any supersymmetric Janus solutions within $SO(3)_{\textrm{diag}}$ sector obtained by setting $\phi_0=0$ and $\phi_1=\phi_2$. On the other hand, for $\phi_1=0$, we recover the BPS equations studied in \cite{6D_Janus}. Finally, we note that the solutions preserve half of the $N=(1,1)$ supersymmetry or eight supercharges since there is only one projection condition \eqref{gamma_r_proj} imposed on the Killing spinors. 
\subsection{Numerical Janus solutions}
From the previous analysis, we find a set of BPS equations for obtaining Janus solutions. Before finding the solutions, it is convenient to collect all the relevant equations here
\begin{eqnarray}
& &\phi'_1=\eta \sqrt{M_1^2+M_2^2},\\
& &\phi'_0=-\eta\frac{(M_0M_1+M_2M_3)}{\cosh\phi_2\sqrt{M_1^2+M_2^2}},\\
& &\phi'_2=\eta\frac{M_1M_3-M_0M_2}{\sqrt{M_1^2+M_2^2}},\\
& &\sigma'=\eta\frac{2(N_0M_1-N_3M_2)}{\sqrt{M_1^2+M_2^2}},\\
& &f'=-\eta\frac{2(S_0M_1+S_3M_2)}{\sqrt{M_1^2+M_2^2}}
\end{eqnarray}
for $\eta=\pm 1$ as in \cite{6D_Janus}, together with two algebraic constraints
\begin{eqnarray}
& &N_0M_2=-N_3M_1,\\
& &e^{-2f}=4(S_0^2+S_3^2)-\frac{4(S_0M_1+S_3M_2)^2}{M_1^2+M_2^2}\, .
\end{eqnarray}
The two values of $\eta$ correspond to two branches of solutions that need to be glued together to obtain smooth solutions \cite{6D_Janus}. These BPS equations are too complicated to be solved analytically. Accordingly, we will look for numerical Janus solutions. We will restrict ourselves to regular Janus solutions interpolating between two $AdS_6$ vacua on both sides. In the numerical search for solutions, there are also singular solutions with scalars or the warped factor $f(r)$ diverge on one or both sides of the interfaces. However, we will not consider these solutions in this paper. 
\\
\indent Before giving examples of numerical Janus solutions, we first look at asymptotic behaviors of scalar fields near the $AdS_6$ vacua. The scalar masses at both critical points have been given in \cite{F4_flow}. For convenience, we repeat this result here and also include the values of scaling dimensions of the dual operators in the $N=2$ five-dimensional SCFTs via the relation 
 \begin{equation}
 \Delta(\Delta-5)=m^2L^2\, .
 \end{equation}
These values are given in table \ref{table1} and \ref{table2} for the $SO(4)$ and $SO(3)_{\textrm{diag}}$ $AdS_6$ vacua, respectively. In table \ref{table2}, we have distinguished between the vector and adjoint representations of $SO(3)_{\textrm{diag}}$ for clarity. The massless adjoint scalars correspond to Goldstone bosons of the symmetry breaking $SO(4)\rightarrow SO(3)_{\textrm{diag}}$.  
 \begin{table}[h]
\centering
\begin{tabular}{|c|c|c|}
  \hline
scalars & $m^2L^2\phantom{\frac{1}{2}}$ & $\Delta$  \\ \hline
$(\mathbf{1},\mathbf{1})$ &   $-6$ &  $3$  \\
$(\mathbf{1},\mathbf{3})$  & $-4$ &  $4$  \\
$(\mathbf{3},\mathbf{3})$  &  $-6$ &  $3$  \\
  \hline
\end{tabular}
\caption{Scalar masses at the $SO(4)$ supersymmetric $AdS_6$ vacuum and the
corresponding dimensions of the dual operators.}\label{table1}
\end{table}

\begin{figure}
         \centering
               \begin{subfigure}[b]{0.45\textwidth}
                 \includegraphics[width=\textwidth]{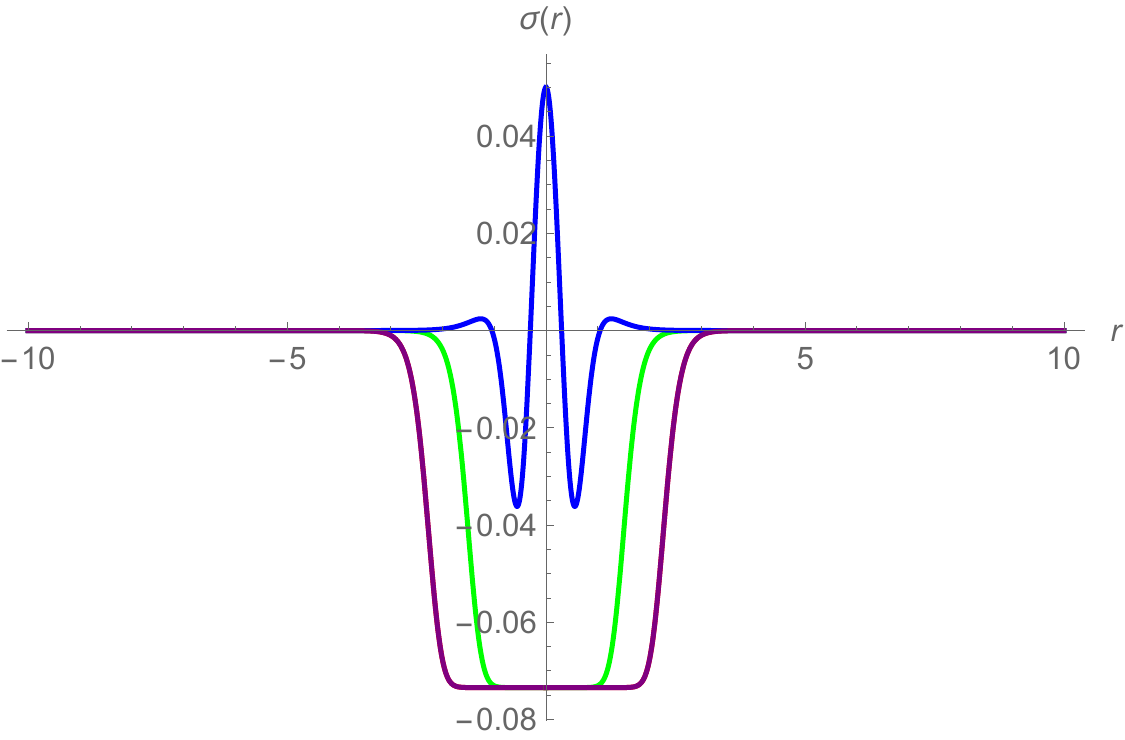}
                 \caption{Solutions for $\sigma(r)$}
         \end{subfigure}
         \begin{subfigure}[b]{0.45\textwidth}
                 \includegraphics[width=\textwidth]{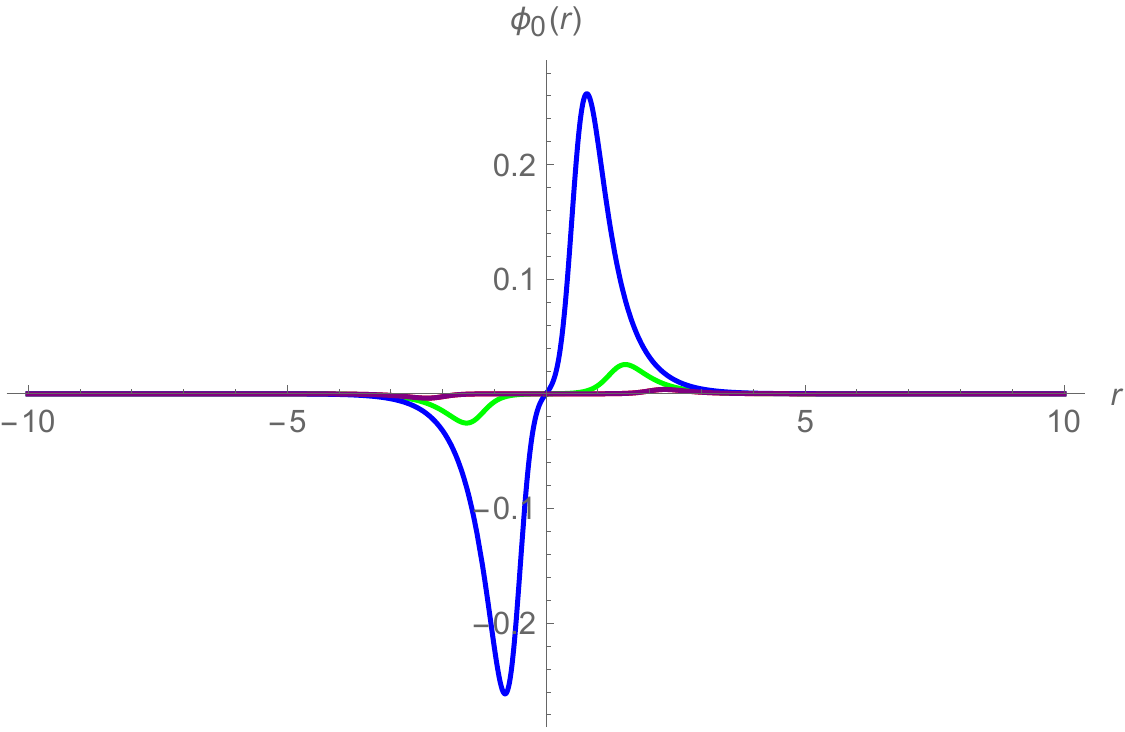}
                 \caption{Solutions for $\phi_0(r)$}
         \end{subfigure}\\
          \begin{subfigure}[b]{0.45\textwidth}
                 \includegraphics[width=\textwidth]{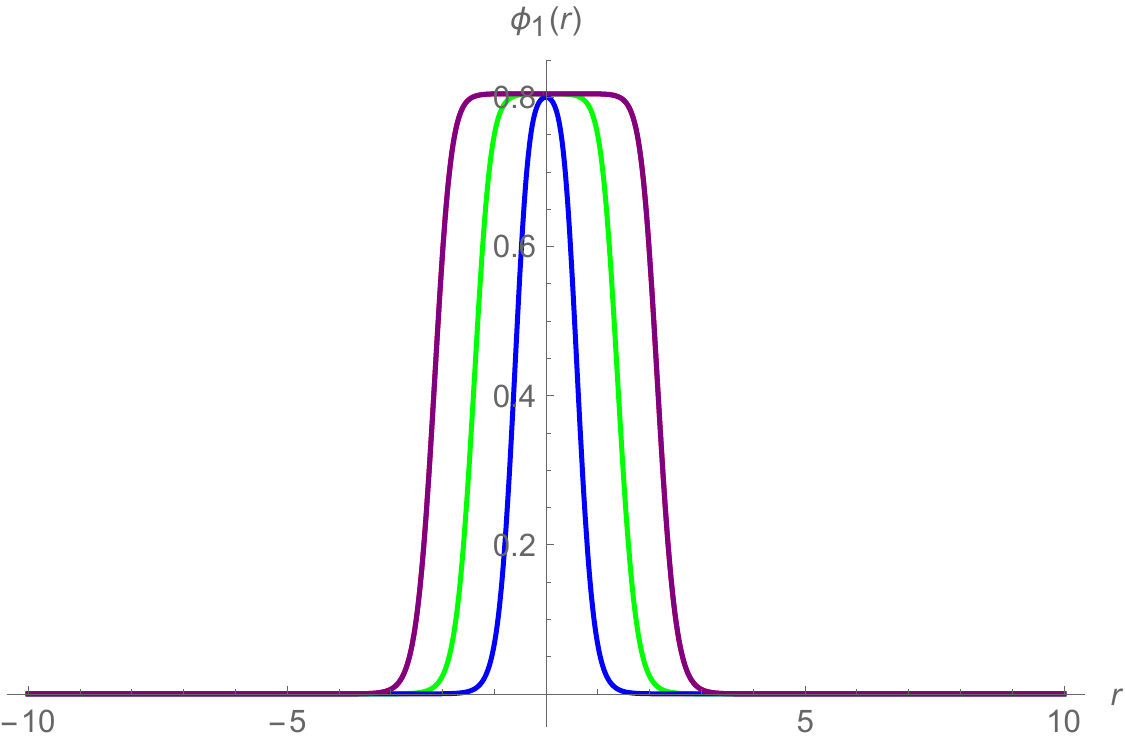}
                 \caption{Solutions for $\phi_1(r)$}
         \end{subfigure}
          \begin{subfigure}[b]{0.45\textwidth}
                 \includegraphics[width=\textwidth]{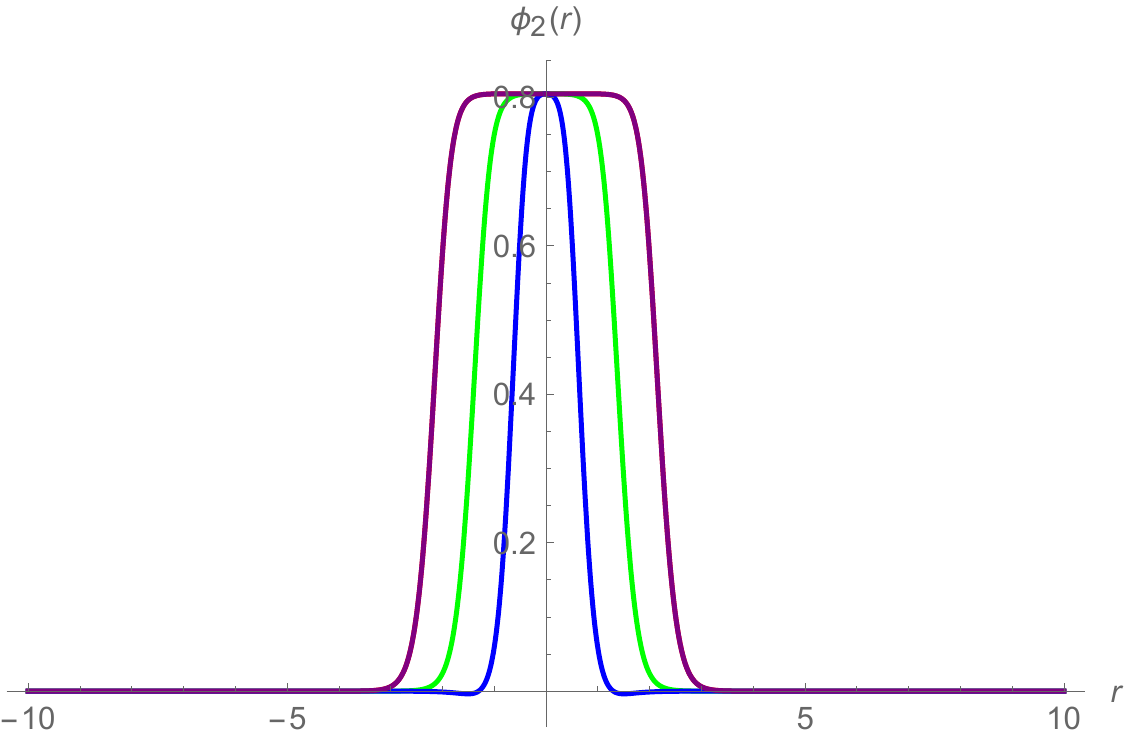}
                 \caption{Solutions for $\phi_2(r)$}
         \end{subfigure}\\
         \begin{subfigure}[b]{0.45\textwidth}
                 \includegraphics[width=\textwidth]{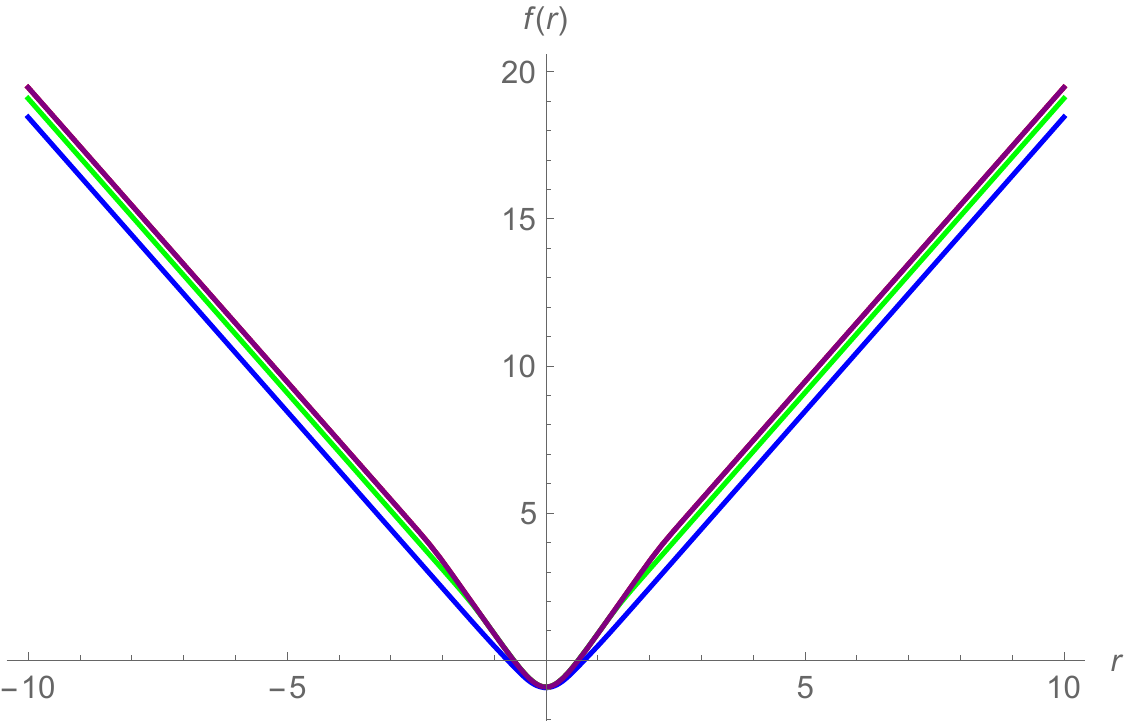}
                 \caption{Solutions for $f(r)$}
         \end{subfigure}
          \begin{subfigure}[b]{0.45\textwidth}
                 \includegraphics[width=\textwidth]{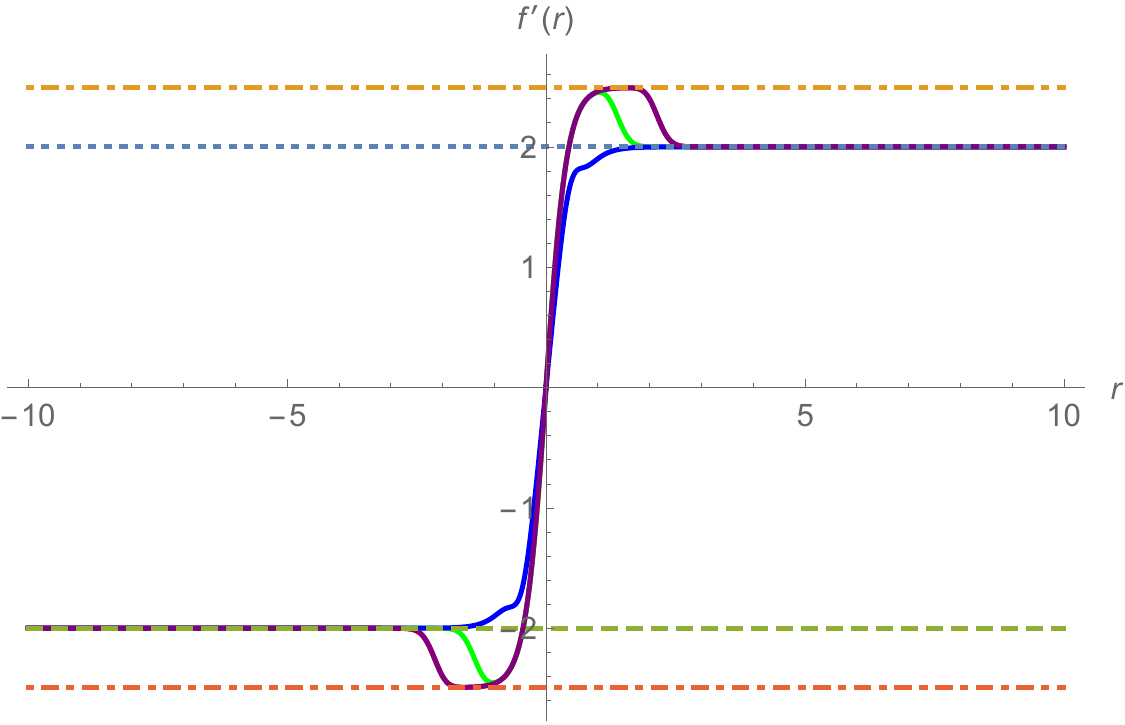}
                 \caption{Solutions for $f'(r)$}
         \end{subfigure}
\caption{Supersymmetric Janus solutions preserving eight supercharges and $SO(2)_{\textrm{diag}}$ symmetry from matter-coupled $F(4)$ gauged supergravity with $SO(3)\times SO(3)$ gauge group for $m=1$, $g_1=3m$ and $g_2=\frac{3}{2}g_1$. The blue line represents a solution interpolating between $SO(4)$ $AdS_6$ vacua while the green and purple lines are solutions interpolating between $SO(3)_{\textrm{diag}}$ critical points on both sides generated by RG flows from the $SO(4)$ critical point.}\label{fig1}
 \end{figure}

\begin{figure}
         \centering
               \begin{subfigure}[b]{0.45\textwidth}
                 \includegraphics[width=\textwidth]{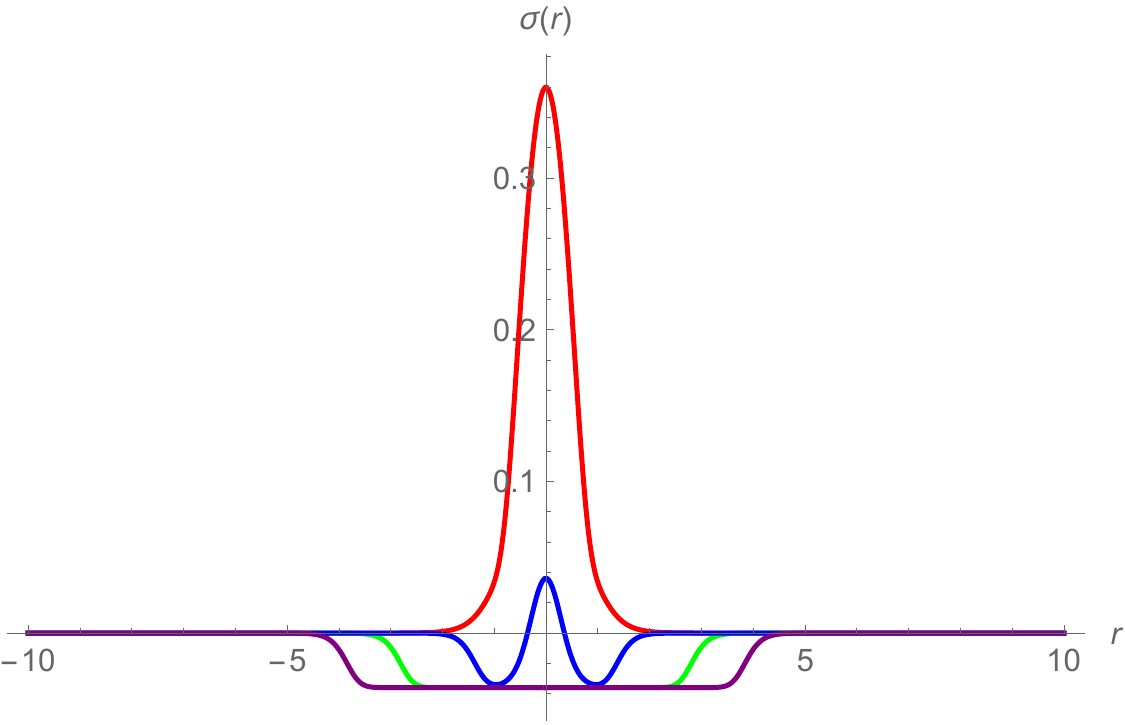}
                 \caption{Solutions for $\sigma(r)$}
         \end{subfigure}
         \begin{subfigure}[b]{0.45\textwidth}
                 \includegraphics[width=\textwidth]{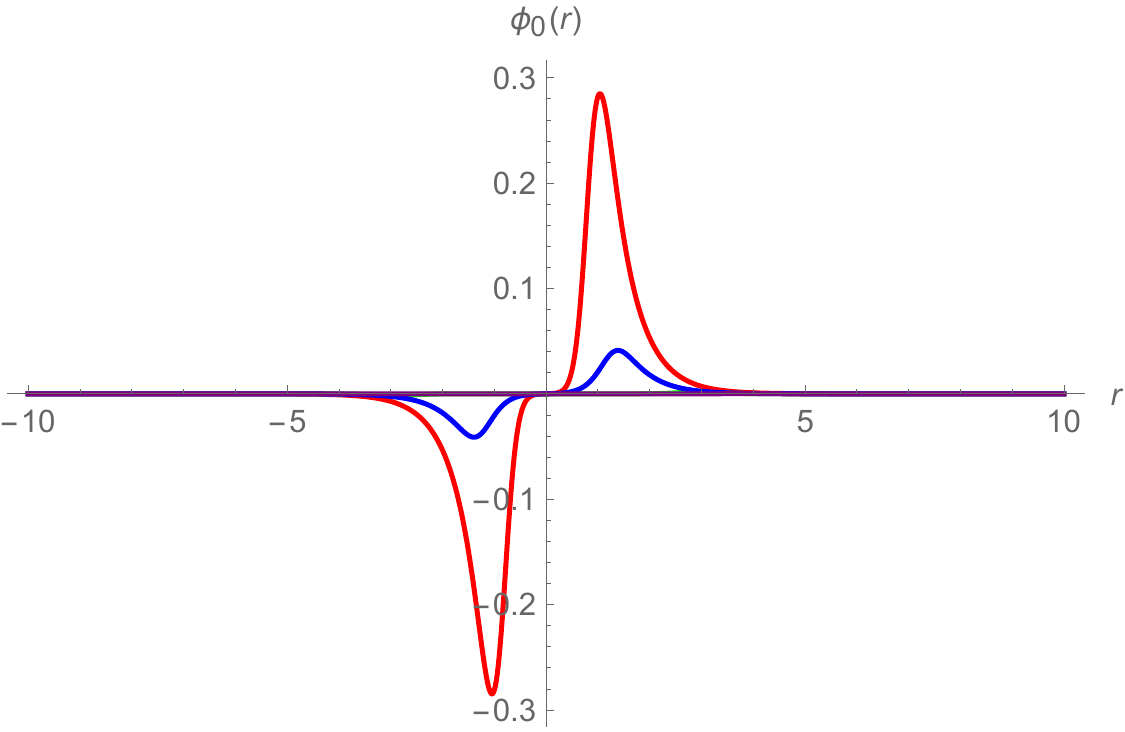}
                 \caption{Solutions for $\phi_0(r)$}
         \end{subfigure}\\
          \begin{subfigure}[b]{0.45\textwidth}
                 \includegraphics[width=\textwidth]{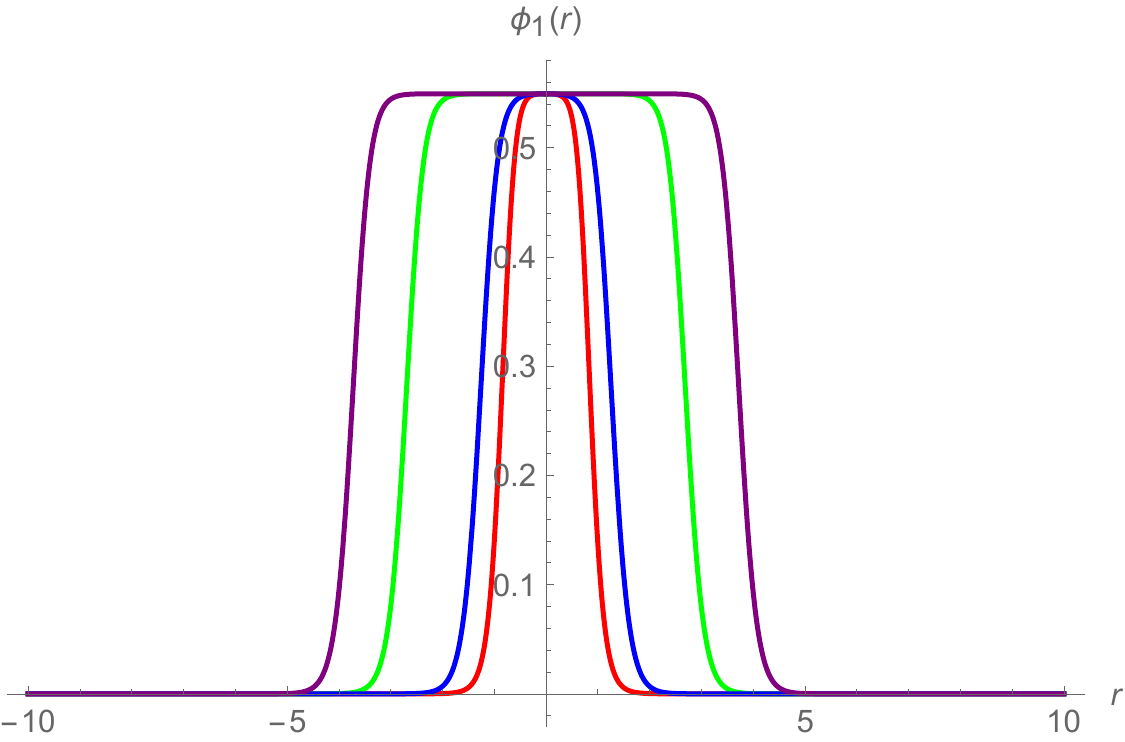}
                 \caption{Solutions for $\phi_1(r)$}
         \end{subfigure}
          \begin{subfigure}[b]{0.45\textwidth}
                 \includegraphics[width=\textwidth]{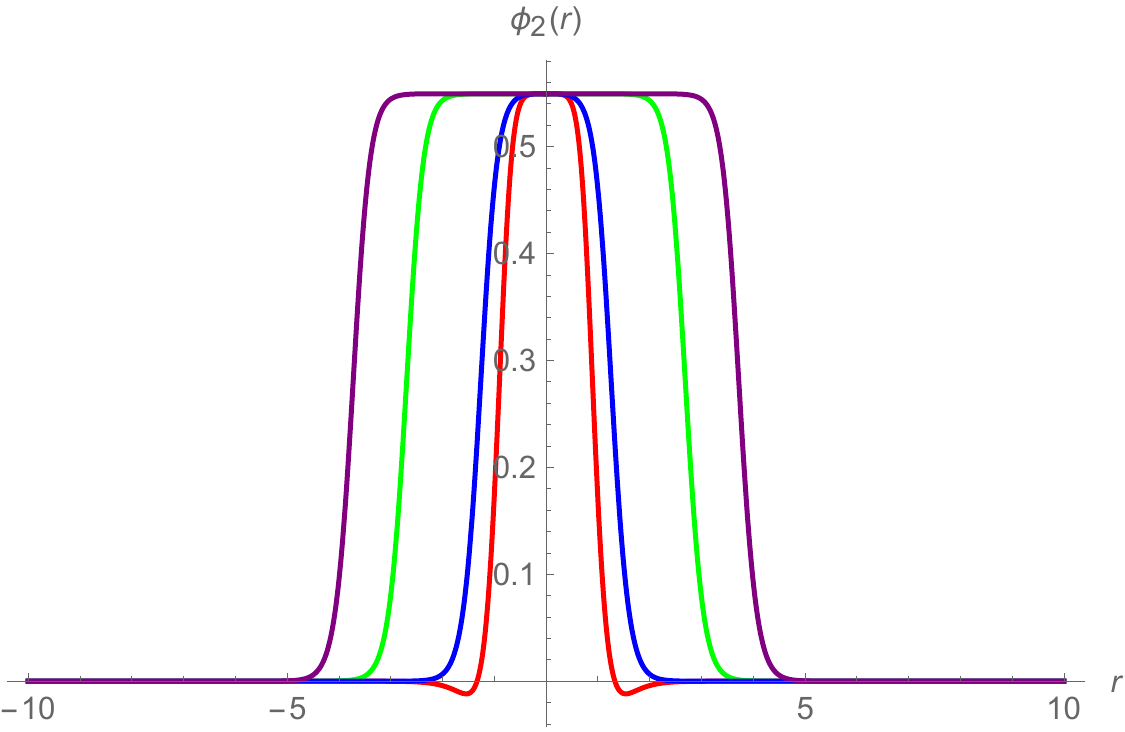}
                 \caption{Solutions for $\phi_2(r)$}
         \end{subfigure}\\
         \begin{subfigure}[b]{0.45\textwidth}
                 \includegraphics[width=\textwidth]{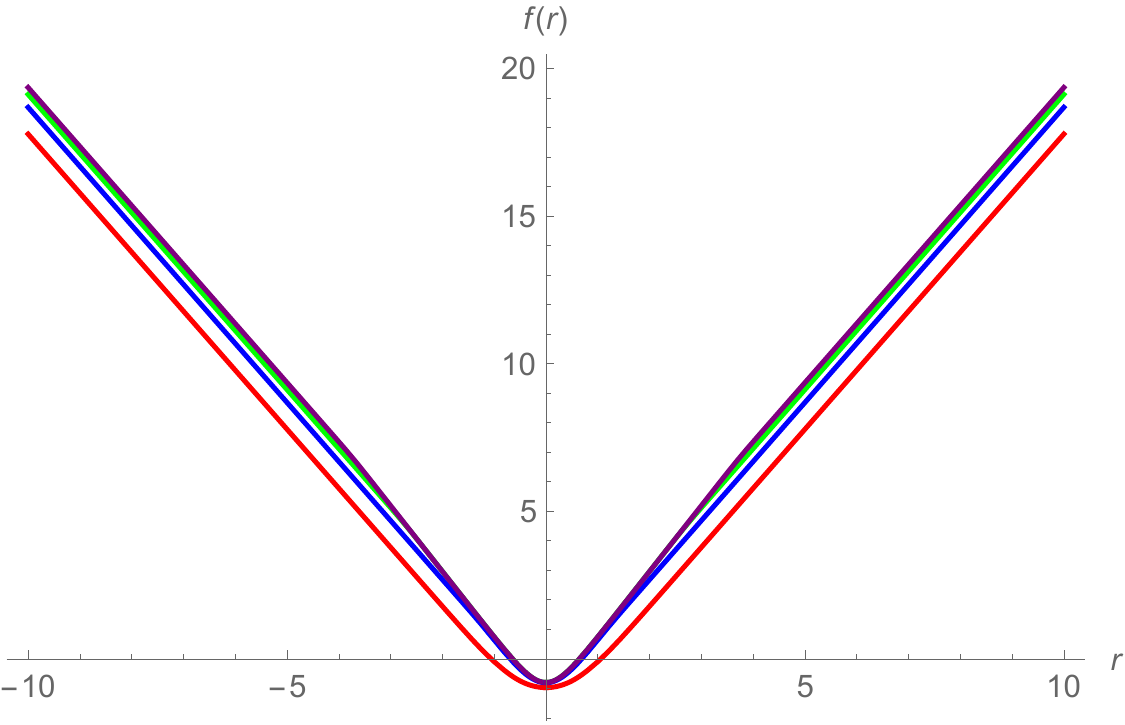}
                 \caption{Solutions for $f(r)$}
         \end{subfigure}
          \begin{subfigure}[b]{0.45\textwidth}
                 \includegraphics[width=\textwidth]{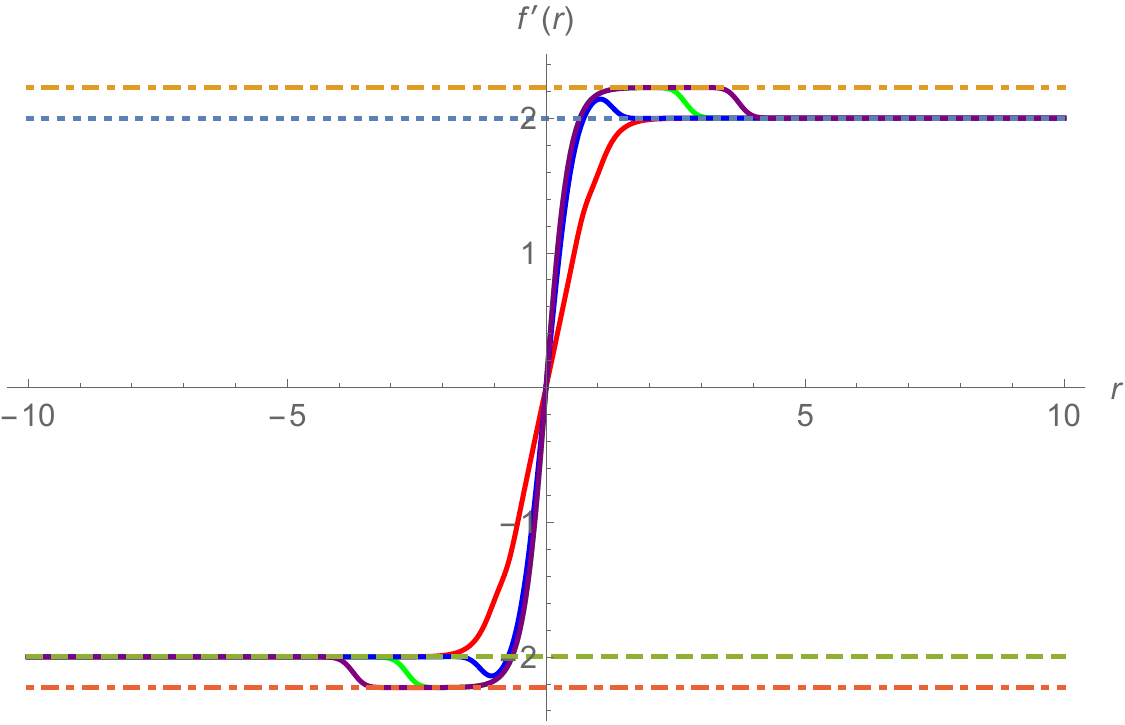}
                 \caption{Solutions for $f'(r)$}
         \end{subfigure}
\caption{Supersymmetric Janus solutions preserving eight supercharges and $SO(2)_{\textrm{diag}}$ symmetry from matter-coupled $F(4)$ gauged supergravity with $SO(3)\times SO(3)$ gauge group for $m=1$, $g_1=3m$ and $g_2=2g_1$. The red line represents a solution interpolating between $SO(4)$ $AdS_6$ vacua while the blue, green and purple lines are solutions interpolating between $SO(3)_{\textrm{diag}}$ critical points on both sides generated by RG flows from the $SO(4)$ critical point.}\label{fig2}
 \end{figure}

\begin{table}[h]
\centering
\begin{tabular}{|c|c|c|}
  \hline
scalars & $m^2L^2\phantom{\frac{1}{2}}$ & $\Delta$  \\ \hline
 $\mathbf{1}$ &  $-6$ &  $3$  \\
$\mathbf{1}$  & $24$ &  $8$  \\
 $\mathbf{3}$  & $14$ &  $7$  \\
$\mathbf{3}_{\textrm{Adj}}$ & $0$ &  $5$  \\
$\mathbf{5}$ & $6$ &  $6$  \\
  \hline
\end{tabular}
\caption{Scalar masses at the $SO(3)_{\textrm{diag}}$ supersymmetric $AdS_6$ vacuum and the
corresponding dimensions of the dual operators.}\label{table2}
\end{table}
\indent At the $SO(4)$ $AdS_6$ vacuum, the scalar $\sigma$ corresponds to $(\mathbf{1},\mathbf{1})$ while $\phi_0$ and $(\phi_1,\phi_2)$ are parts of $(\mathbf{1},\mathbf{3})$ and $(\mathbf{3},\mathbf{3})$ scalars, respectively. From table \ref{table1}, all of these scalars are dual to relevant operators, so the $SO(4)$ $AdS_6$ vacuum is an attractive critical point at large $r$. This can also be seen from the linearied BPS equations near the asymptotic $AdS_6$ geometry which for $g_1=3m$ give
\begin{eqnarray}
\sigma\sim \phi_1\sim \phi_2 \sim e^{-6mr}\sim z^3\qquad\textrm{and}\qquad \phi_0\sim e^{-\frac{r}{L}}\sim z\, .
\end{eqnarray}
As in \cite{3D_Janus2} and \cite{3D_Janus3}, we have introduced a new radial coordinate $z$ related to $r$ via $z=e^{-\frac{r}{L}}$ for $L=\frac{1}{2m}$. In this coordinate, the $AdS_6$ boundary at $r\rightarrow \infty$ is given by $z\rightarrow 0$. The solutions generally behave as $z^{5-\Delta}$ or $z^\Delta$ with the coefficients corresponding respectively to position-dependent (along the direction transverse to the interface) sources and expectation values \cite{6D_Janus}. Therefore, we see that the source of a dimension $4$ operator dual to $\phi_0$ is turned on in the presence of vacuum expectation values for operators of dimension $3$ dual to $\sigma$, $\phi_1$ and $\phi_2$. 
\\  
\indent On the other hand, at the $SO(3)_{\textrm{diag}}$ $AdS_6$ vacuum, the dilaton $\sigma$ (the first singlet in table \ref{table2}) is still dual to a relevant operator of dimension $\Delta=3$ while, in this case, $\phi_0$ and $(\phi_1,\phi_2)$, which are respectively parts of $\mathbf{3}$ and $\mathbf{1}$ (the second singlet) together with $\mathbf{5}$ scalars, are dual to irrelevant operators of dimensions $\Delta=6,7,8$. The linearized BPS equations near this $AdS_6$ critical point give
\begin{eqnarray}
& &\sigma\sim e^{-\frac{3r}{L}}\sim z^3,\qquad \phi_0\sim e^{\frac{2r}{L}}\sim z^{-2},\nonumber \\
& & \phi_1+\phi_2\sim e^{\frac{3r}{L}}\sim z^{-3},\qquad 2\phi_1-\phi_2\sim e^{-\frac{6r}{L}}\sim z^6 \label{AdS6_1_asymp}
\end{eqnarray}                           
with $L=\frac{1}{2m}\left(1-\frac{9m^2}{g^2_2}\right)^{\frac{3}{4}}$. From table \ref{table2}, $\sigma$ and $\phi_0$ are dual to operators of dimensions $3$ and $7$. $\phi_1+\phi_2$ corresponding to the second $SO(3)_{\textrm{diag}}$ singlet in the table is dual to a dimension $8$ operator while $2\phi_1-\phi_2$ in $\mathbf{5}$ representation of $SO(3)_{\textrm{diag}}$ is dual to a dimension $6$ operator. 
\\
\indent From equation \eqref{AdS6_1_asymp}, there are diverging scalars when the solution approaches the $SO(3)_{\textrm{diag}}$ critical point. The $SO(3)_{\textrm{diag}}$ $AdS_6$ vacuum is then a repulsive critical point, and we need to fine tune the boundary conditions to obtain solutions connecting this critical point, see also the results in \cite{3D_Janus2,3D_Janus3} and \cite{warner_Janus} for similar solutions within three- and four-dimensional gauged supergravities. However, the process of fine tuning in the numerical analysis turns out to be rather difficult (the solutions in \cite{warner_Janus,3D_Janus2,3D_Janus3} only involve two active scalars). 
\\
\indent To find numerical solutions, we first choose particular values of various parameters as follows
\begin{equation}
g_1=3m\qquad \textrm{and}\qquad m=1\, .
\end{equation}
The value of $g_1=3m$ can be chosen by shifting the value of the dilaton $\sigma$ at the $SO(4)$ $AdS_6$ vacuum to zero. The value of $m$ is related to the corresponding $AdS_6$ radius at this vacuum via $L=\frac{1}{2m}$, so the value of $m$ can be chosen by setting this $AdS_6$ radius at a particular value. On the other hand, the value of $g_2$ is only constrained to be $g_2>g_1$ in order for the $SO(3)_{\textrm{diag}}$ $AdS_6$ vacuum to exist. Following the procedure describe in \cite{6D_Janus}, we find examples of Janus solutions shown in figures \ref{fig1} and \ref{fig2} for two different values of $g_2=\frac{3}{2}g_1$ and $g_2=2g_1$. 
\\
\indent As in the cases of similar solutions in other dimensions, there are solutions interpolating between the $SO(4)$ $AdS_6$ vacua on both sides of the interfaces shown by the blue line in figure \ref{fig1} and the red line in figure \ref{fig2}. By fine-tuning the boundary condition at the turning point, we can find solutions that flow from the $SO(4)$ critical point to the $SO(3)_{\textrm{diag}}$ $AdS_6$ vacua before reaching the turning point and flow back to the $SO(4)$ critical point again. These solutions are shown by green and purple lines in figure \ref{fig1} and blue, green and purple lines in figure \ref{fig2}. The $SO(3)_{\textrm{diag}}$ phases appearing on both sides of the interfaces are generated by the usual holographic RG flows from the $SO(4)$ phases. In the solutions for $f'(r)$, we have included the values of $\frac{1}{L}$ with $L$ being the $AdS_6$ radii at different vacua for clarity. By tuning the turning point to approach the $SO(3)_{\textrm{diag}}$ critical point, we can make the solutions stay at this critical point longer before flowing to the $SO(4)$ critical point, see an example shown by the purple line in figure \ref{fig2}. Following \cite{warner_Janus}, we interpret these solutions as describing Janus interfaces between $SO(3)_{\textrm{diag}}$ $AdS_6$ vacua at both sides of the interfaces.

\begin{figure}
         \centering
               \begin{subfigure}[b]{0.45\textwidth}
                 \includegraphics[width=\textwidth]{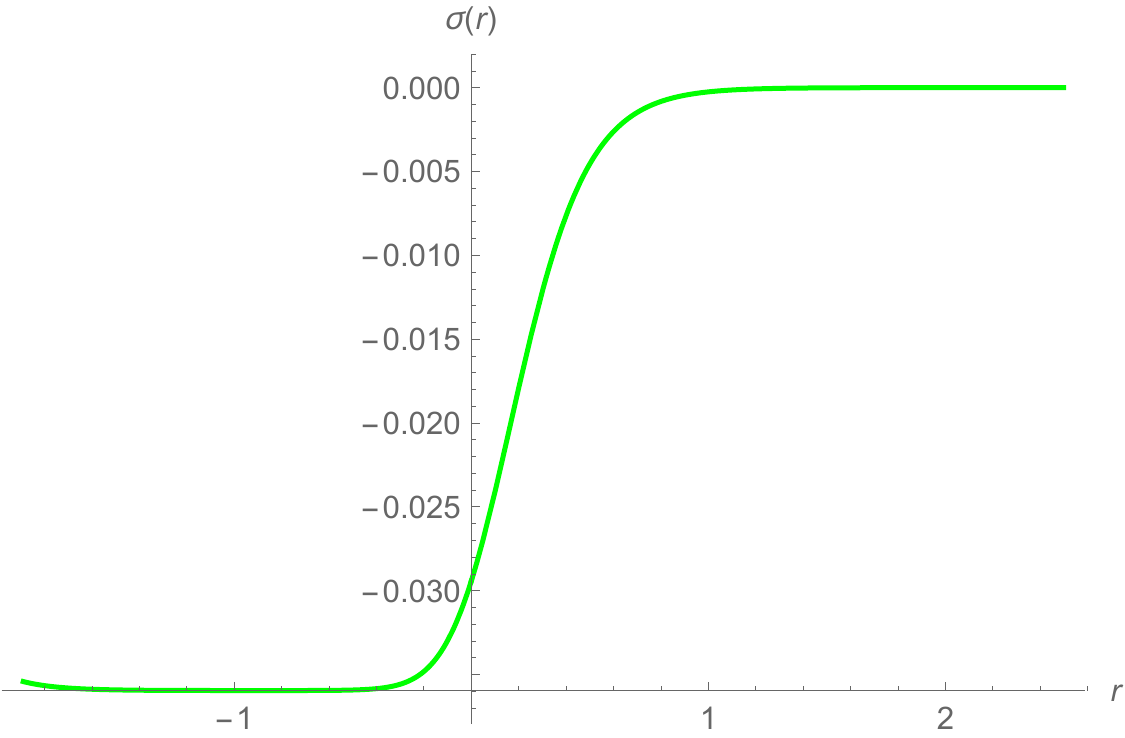}
                 \caption{Solution for $\sigma(r)$}
         \end{subfigure}
         \begin{subfigure}[b]{0.45\textwidth}
                 \includegraphics[width=\textwidth]{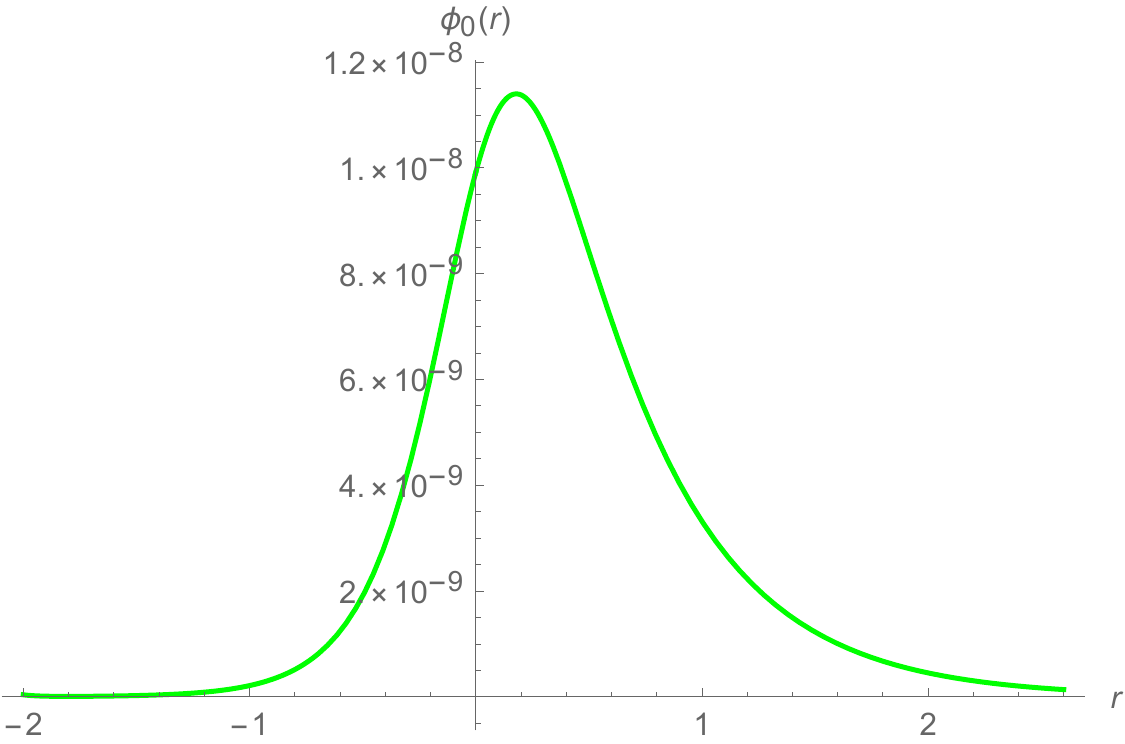}
                 \caption{Solution for $\phi_0(r)$}
         \end{subfigure}\\
          \begin{subfigure}[b]{0.45\textwidth}
                 \includegraphics[width=\textwidth]{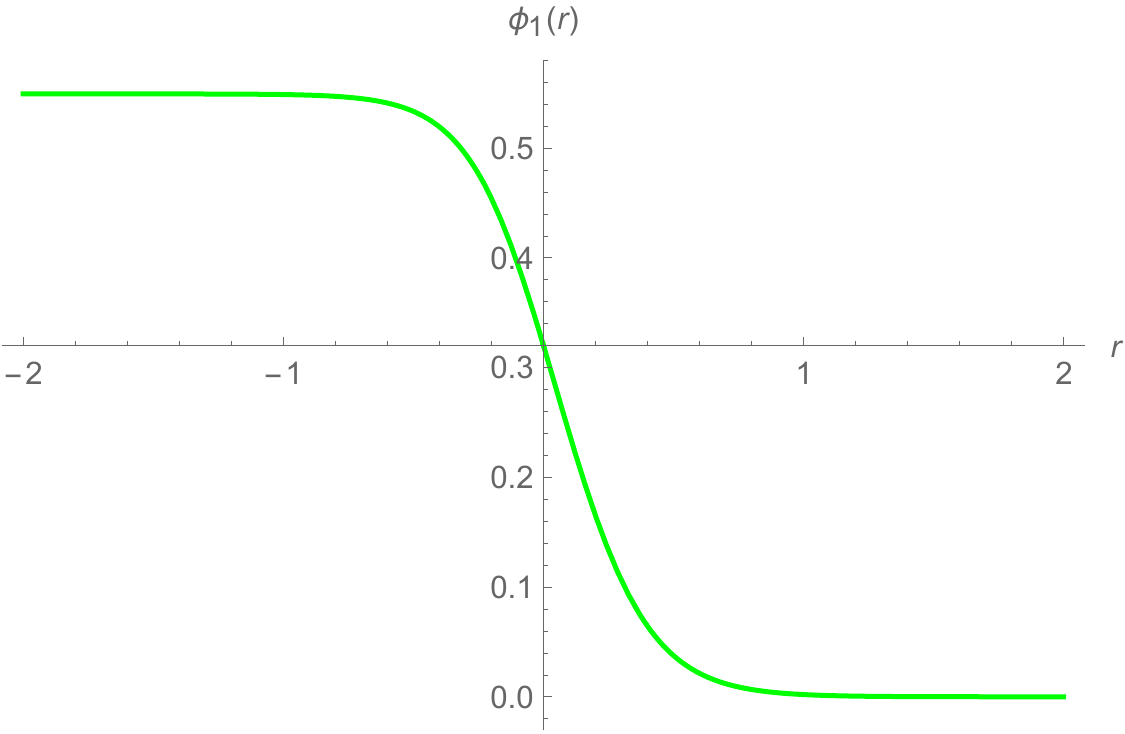}
                 \caption{Solution for $\phi_1(r)$}
         \end{subfigure}
          \begin{subfigure}[b]{0.45\textwidth}
                 \includegraphics[width=\textwidth]{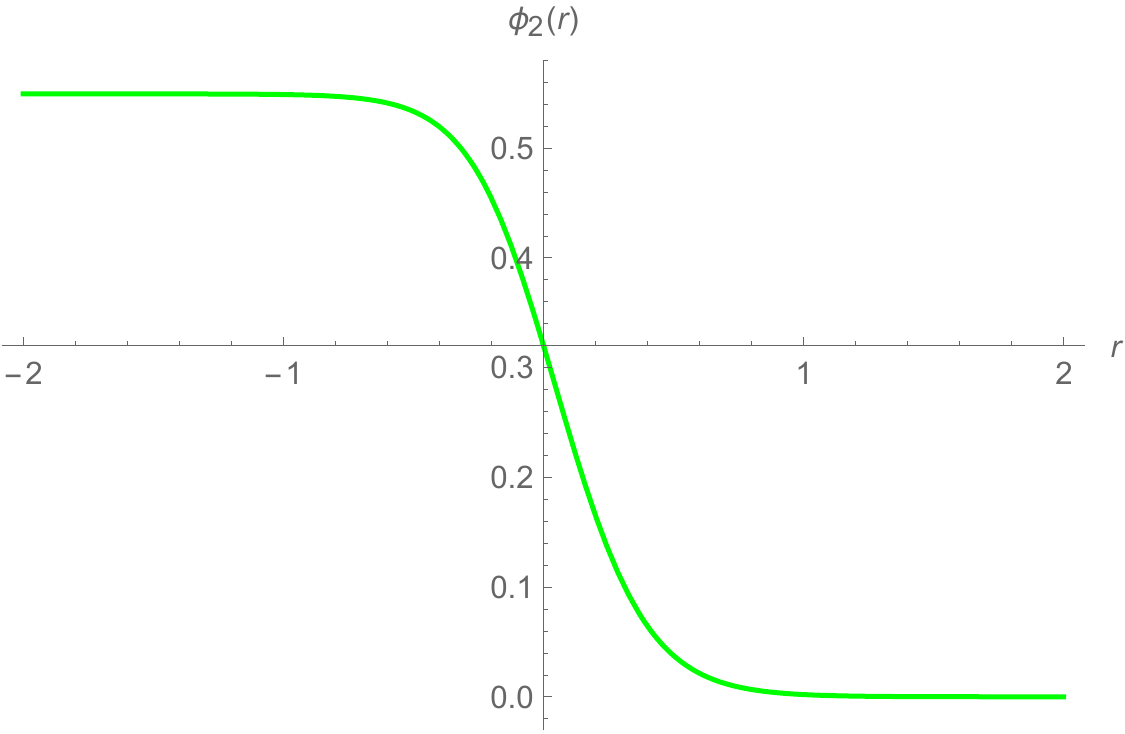}
                 \caption{Solution for $\phi_2(r)$}
         \end{subfigure}\\
         \begin{subfigure}[b]{0.45\textwidth}
                 \includegraphics[width=\textwidth]{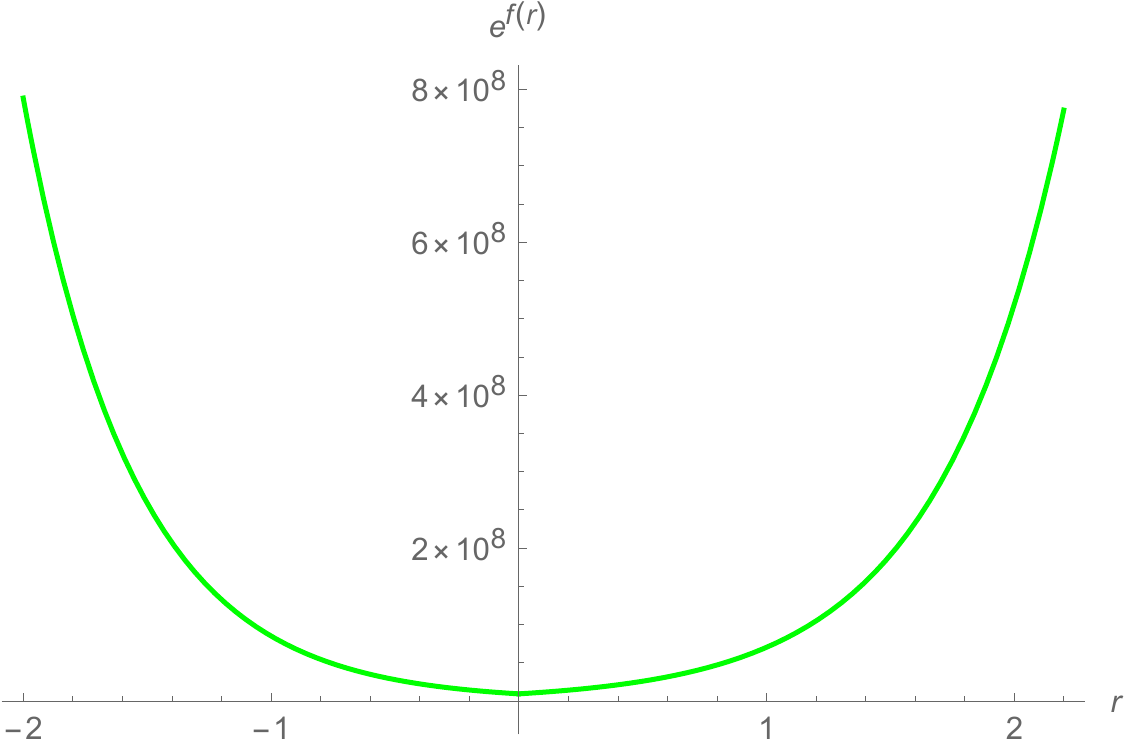}
                 \caption{Solution for $e^{f(r)}$}
         \end{subfigure}
          \begin{subfigure}[b]{0.45\textwidth}
                 \includegraphics[width=\textwidth]{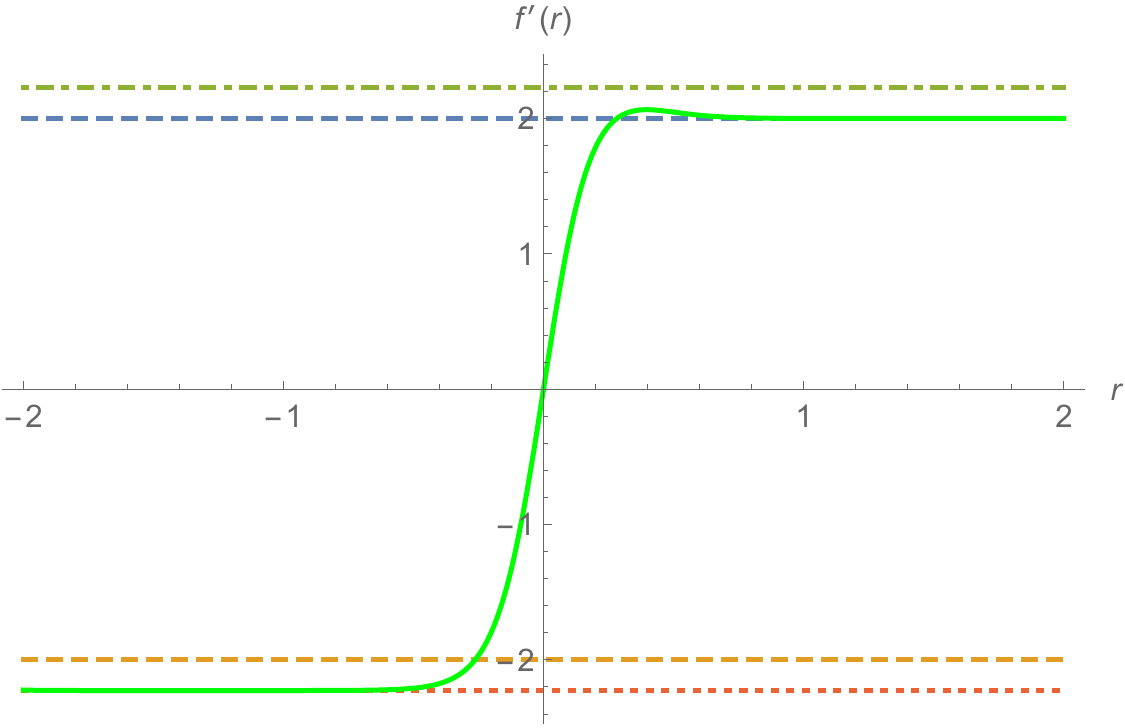}
                 \caption{Solution for $f'(r)$}
         \end{subfigure}
\caption{A supersymmetric Janus solution interpolating between the $SO(3)_{\textrm{diag}}$ $AdS_6$ vacuum on one side and the $SO(3)\times SO(3)$ $AdS_6$ critical point on the other side for $m=1$, $g_1=3m$ and $g_2=2g_1$.}\label{fig3}
 \end{figure}

\indent After an intensive numerical search with a large number of different boundary conditions, we find an example of RG-flow interfaces shown in figure \ref{fig3}. In this case, the solution interpolates between the $SO(3)_{\textrm{diag}}$ critical point on one side and the $SO(4)$ critical point on the other side. Accordingly, we interpret the solution as describing a Janus interface between the $SO(3)_{\textrm{diag}}$ critical point on one side and the $SO(4)$ critical point on the other side of the interface. This is an example of RG-flow interfaces interpolating between different SCFTs on each side of the interfaces.

\section{Conclusions}\label{conclusion}
We have constructed new supersymmetric Janus solutions from matter-coupled $F(4)$ gauged supergravity in six dimensions with $SO(3)\times SO(3)$ gauge group. The solutions are $SO(2)_{\textrm{diag}}$ invariant and preserve eight supercharges. We have found numerical solutions which are holographically dual to conformal interfaces between $SO(4)$ $AdS_6$ vacua and between $SO(3)_{\textrm{diag}}$ $AdS_6$ vacua. The first class of solutions describe conformal interfaces with the $N=2$ $SO(3)\times SO(3)$ SCFTs on both sides deformed by a source of a relevant operator of dimension $4$ and position-dependent expectation values for operators of dimension $3$. The second type of solutions provides a holographic description of conformal interfaces between $N=2$ $SO(3)_{\textrm{diag}}$ SCFTs with irrelevant sources of dimensions $7$ and $8$ in the presence of expectation values for relevant and irrelevant operators of dimensions $3$ and $6$. By fine-tuning the boundary conditions to approach the repulsive $SO(3)_{\textrm{diag}}$ $AdS_6$ critical point, we have also given an evidence for the existence of RG-flow interfaces interpolating between $SO(4)$ and $SO(3)_{\textrm{daig}}$ $AdS_6$ vacua on each side. 
\\
\indent It would be interesting to construct conformal interfaces dual to the solutions presented here in the framework of five-dimensional $N=2$ SCFTs. Using the Janus solutions in this paper to holographically study mass deformations of $N=2$ SCFTs along the line of \cite{mass_deform_5D_SCFT} is worth considering. Furthermore, finding possible consistent truncations of string/M-theory to six-dimensional gauged supergravity considered here is also interesting and could provide a means of uplifting the Janus solutions to ten or eleven dimensions. Although the results of \cite{Henning_Malek_AdS7_6} do not allow for uplifting the $SO(3)_{\textrm{diag}}$ $AdS_6$ vacuum and the associated Janus solutions to type IIB theory, by including the gauge group from vector fields on D-branes, it might be possible to obtained this $AdS_6$ vacuum by making two D5-branes end on a D7-brane \cite{AdS6_IIB1,AdS6_IIB2,AdS6_IIB3}. On the other hand, performing a similar analysis to find Janus solutions in six-dimensional gauged supergravity with presently known higher-dimensional origins classified in \cite{Henning_Malek_AdS7_6} is also of particular interest. We leave these interesting issues for future works.  
\begin{acknowledgments}
This work is funded by National Research Council of Thailand (NRCT) and Chulalongkorn University under grant N42A650263.
\end{acknowledgments}
\appendix
\section{Bosonic field equations}
In this appendix, we collect all the scalar and Einstein field equations within the truncation to $SO(2)_{\textrm{diag}}$ singlet scalars. These equations read
\begin{eqnarray}
\sigma''+5f'\sigma'-\frac{1}{2}\frac{\pd V}{\pd \sigma}&=&0,\\
\cosh^2\phi_2\phi''_0+5\cosh^2\phi_2f'\phi'_0+2\cosh\phi_2\sinh\phi_2\phi'_0\phi'_2-2\frac{\pd V}{\pd \phi_0}&=&0,\\
\phi''_1+5f'\phi'_1-\frac{\pd V}{\pd\phi_1}&=&0,\\
\phi''_2+5f'\phi'_2-\frac{1}{2}\sinh2\phi_2{\phi'_0}^2-2\frac{\pd V}{\pd \phi_2}&=&0,\\
4f''+10{f'}^2+6e^{-2f}+2{\sigma'}^2+\frac{1}{2}\cosh^2\phi_2{\phi'_0}^2+{\phi'_1}^2+\frac{1}{2}{\phi'_2}^2 +2V&=&0,\\
10{f'}^2+10e^{-2f}-2{\sigma'}^2-\frac{1}{2}\cosh^2\phi_2{\phi'_0}^2-{\phi'_1}^2-\frac{1}{2}{\phi'_2}^2+2V&=&0
\end{eqnarray}
with the scalar potential given by 
\begin{eqnarray}
V&=&g_1^2e^{2\sigma}\cosh^2\phi_1(\cosh2\phi_2\sinh^2\phi_1-1)-4g_1me^{-2\sigma}\cosh\phi_0\cosh^2\phi_1\cosh\phi_2\nonumber \\
& &+m^2e^{-6\sigma}(\cosh^2\phi_0+\cosh2\phi_2\sinh^2\phi_0)+4g_2me^{-2\sigma}\sinh^2\phi_1\sinh\phi_2\nonumber \\
& &-\frac{1}{2}g_1g_2e^{2\sigma}\cosh\phi_0\sinh^2\phi_1\sinh2\phi_2+\frac{1}{8}g_2^2e^{2\sigma}\sinh^2\phi_1\left[\cosh2(\phi_0-\phi_1) \right.\nonumber \\
& &+2\cosh2\phi_0-2\cosh2\phi_1 -10+\cosh2(\phi_0+\phi_1)\nonumber \\
& &\left.+8\cosh^2\phi_0\cosh^2\phi_1\cosh2\phi_2\right].
\end{eqnarray}

\end{document}